\documentclass[a4paper, 12pt, oneside]{article}
\usepackage[cp1251]{inputenc}
\usepackage[english, russian]{babel}
\usepackage{ graphics,amsfonts, amsmath,bm,amssymb, array}
\usepackage[dvips]{graphicx}
\renewcommand{\Im}{\mathop{\rm Im\,}}
\makeatother {

\begin{document}
\thispagestyle{empty} \large
\renewcommand{\abstractname}{}

\renewcommand{\refname}{\begin{center} \bf References\end{center}}

 \begin{center}
\bf Transverse electric conductivity in quantum collisional
plasma in Mermin approach
\end{center}\medskip
\begin{center}
  \bf A. V. Latyshev\footnote{$avlatyshev@mail.ru$} and
  A. A. Yushkanov\footnote{$yushkanov@inbox.ru$}
\end{center}\medskip

\begin{center}
{\it Faculty of Physics and Mathematics,\\ Moscow State Regional
University, 105005,\\ Moscow, Radio str., 10-A}
\end{center}\medskip

\begin{abstract}
Formulas for transversal electric conductivity and dielectric
permeability of qu\-an\-tum collisional plasma are deduced.
The kinetic equation for a density matrix in relaxation approaching in
momentum space is used.
It is shown, that when Planck's constant tends to zero,
these deduced formulas pass in classical expressions and when
frequency of electron collision tends to zero (i.e. plasma passes in
collisionless plasma), the deduced formulas pass in deduced earlier
by Lindhard.
It is shown also, that when the wave number tends  to zero, quantum
conductivity passes in the classical one.
Graphic comparison of the deduced conductivity  with Lindhard's
conductivity and with classical conductivity is carry out.


\end{abstract}

\begin{center}
{\bf  Introduction}
\end{center}
In the known work of Mermin \cite{Mermin} on the basis of the analysis of
nonequilib\-rium density matrix in $\tau $--approximation
has been obtained expression for longitu\-di\-nal dielectric permeability of
quantum collisional plasmas. Earlier in the work of Lindhard \cite{Lin}
has been obtained expression for longitudinal and transversal dielectric
permeability of quantum
collisionless plasmas. By Kliewer and Fuchs \cite{Kliewer} it has been shown,
that direct generalisation of formulas of Lindhard on the case of collisional
plasmas (by replacement $\omega\to \omega+i/\tau$) is incorrectly.
This lack for the longitudinal dielectric
permeability has been eliminated in the work of Mermin \cite{Mermin}.
It is necessary to notice that to present time there is no correct expression for
transversal dielectric permeability in the case of quantum collisional plasmas.
The purpose of our work consists in elimination of this blank.

Properties of electric conductivity and dielectric permeability on
to the formulas deduced by Lindhard \cite{Lin}, were in detail studied in the
monography \cite{Dressel}. In work \cite{Gelder} transversal dielectric
permeability of quantum plasma it was applied in questions of skin effect theory.
Now interest to studying of various properties of the quantum plasmas grows
(see, for example, \cite{Anderson}--\cite{Manf2}).
It is specially necessary
to note work of G. Manfredi \cite{Manf}, devoted to research
of electromagnetic properties of quantum plasma.

\begin{center}
  \bf 1. Kinetic equation for density matrix
\end{center}

Let the vector potential of an electromagnetic field is harmonious, i.e.
it changes as
$
{\bf A}={\bf A}({\bf r})\exp(-i \omega t).
$
We consider the transversal conductivity. So the following condition should
be satisfied
$
{\bf \rm \bf div A}(\mathbf{r},t)=0.
$
Relation between vector potential and intensity of the electric field is given
by the following expression
$
{\bf A}({\bf q})=-({ic}/{\omega}){\bf E}({\bf q}).
$
Equilibrium density matrix has the following form
$$
{\tilde \rho}=\Big[1+\exp\dfrac{H-\mu}{k_BT}\Big]^{-1}.
\eqno{(1.1)}
$$

Here  $T$ is the plasma temperature, $k_B$ is the Bolzmann
constant, $\mu$ is the chemical plasma potential, $H$ is the Hamiltonian.
In linear approximation the Hamiltonian has the following form
$$
H=\dfrac{({\bf p}-({e}/{c}){\bf A})^2}{2m}=
\dfrac{{\bf p}^2}{2m}-\dfrac{e}{2mc}({\bf p}{\bf A}+{\bf A} {\bf
p}).
$$

Here $\mathbf{p}$ is the operator of momentum, $\mathbf {p}=-i\hbar \nabla $,
$e$ and $m$ is the electgon charge and mass, $c$ is the velocity of light.

Therefore we can write the Hamiltonian in the form $H=H_0+H_1$,
where
$$
H_0=\dfrac{{\bf p}^2}{2m},
\qquad H_1=-\dfrac{e}{2mc}({\bf p}{\bf A}+{\bf A} {\bf p}).
$$
We take the kinetic Shr\"{o}dinger equation for density matrix in $\tau$--approximation
$$
i\hbar \dfrac{\partial \rho}{\partial t}=[H,\rho]+
\dfrac{i\hbar}{\tau}({\tilde\rho}-\rho).
\eqno{(1.2)}
$$
Here $\nu=1/\tau$ is the effective collision frequency of electrons,
$\tau$ is the charac\-te\-ris\-tic time between two
consecutive collisions, $\hbar$ is the Planck's constant,
$[H,\rho]=H\rho-\rho H$ is the commutator, $\tilde{\rho}$ is the equilibrium
density matrix.

In linear approximation by external field we shall seek the density matrix
in the form
$$
{\rho}={\tilde \rho}^{(0)}+{\rho}^{(1)}.
\eqno{(1.3)}
$$
Here ${\rho}^{(1)}$ is the correction to equilibrium density matrix,
caused by presence of electromagnetic field,
$\tilde{\rho}^{(0)}$ is the equilibrium density matrix, corresponding to operator $H_0$.

We present equilibrium density matrix $\tilde{\rho}$ in the following
form:
$$
\tilde{\rho}=\tilde{\rho}^{(0)}+\tilde{\rho}^{(1)}.
\eqno{(1.4)}
$$

We consider the commutator $[H, \tilde{\rho}]$. In linear approximation
this commuta\-tor equals to
$$
[H, {\tilde \rho}\,]=[H_0, {\tilde \rho}^{(1)}]+[H_1, {\tilde
\rho}^{(0)}]
\eqno{(1.5)}
$$
and
$$
[H, {\tilde \rho}\,]=0.
\eqno{(1.6)}
$$

For commutators from right side of equality (1.5) we have
$$
\langle\mathbf{k}_1|[H_0, \tilde{\rho}^{(1)}|\mathbf{k}_2]\rangle=
\big(E_{\mathbf{k}_1}-E_{\mathbf{k}_2}\big)\tilde{\rho}^{(1)}
(\mathbf{k}_1-\mathbf{k}_2),
\eqno{(1.7)}
$$
and
$$
\langle\mathbf{k}_1|[H_1, \tilde{\rho}^{(0)}|\mathbf{k}_2]\rangle=
\big(f_{\mathbf{k}_2}-f_{\mathbf{k}_1}\big)\langle\mathbf{k}_1|H_1|
\mathbf{k}_2\rangle=
$$
$$
=\dfrac{e}{2mc}\big(f_{\mathbf{k}_2}-f_{\mathbf{k}_1}\big)
(\mathbf{k}_1+\mathbf{k}_2)\mathbf{A}(\mathbf{k}_1-\mathbf{k}_2),
\eqno{(1.8)}
$$
where
$$
f_{\mathbf{k}}=\Big[1+\exp\dfrac{E_{\mathbf{k}}-\mu}{k_BT}\Big]^{-1},\qquad
E_{\mathbf{k}}=\dfrac{\hbar^2\mathbf{k}^2}{2m}, \qquad
\mathbf{p}=\hbar \mathbf{k}.
$$

From equalities (1.4)--(1.8) it follows that
$$
\tilde{\rho}^{(1)}(\mathbf{k}_1-\mathbf{k}_2)=
-\dfrac{e\hbar}{2mc}\dfrac{f_{\mathbf{k}_1}-
f_{\mathbf{k}_2}}{E_{\mathbf{k}_1}-E_{\mathbf{k}_2}}
(\mathbf{k}_1+\mathbf{k}_2)\mathbf{A}(\mathbf{k}_1-\mathbf{k}_2).
\eqno{(1.9)}
$$

With the help of equalities (1.3)--(1.5) we linearize the equation (1.2).
We obtain the following equation
$$
i\hbar \dfrac{\partial \rho^{(1)}}{\partial t}=[H_0,
\rho^{(1)}]+[H_1, \tilde{\rho}^{(0)}]+i\hbar(\tilde{\rho}^{(1)}-
\rho^{(1)}).
\eqno{(1.10)}
$$

Let's notice that perturbation $\rho^{(1)}\sim \exp(-i\omega t) $,
Then the equation (1.10)  will be transformed to the following form
$$
(\hbar \omega+i\hbar \nu)\rho^{(1)}=[H_0, \rho^{(1)}]+
[H_1, \tilde{\rho}^{(0)}]+i\hbar \nu \tilde{\rho}^{(1)}.
$$
From here we found that
$$
(\hbar \omega+i\hbar \nu)\langle\mathbf{k}_1|\rho^{(1)}|\mathbf{k}_2\rangle=
\langle\mathbf{k}_1|[H_0,\rho^{(1)}]|\mathbf{k}_2\rangle+
$$
$$
+\langle\mathbf{k}_1|[H_1,\tilde{\rho}^{(0)}]|\mathbf{k}_2\rangle+i\hbar \nu
\langle\mathbf{k}_1|\tilde{\rho}^{(0)}|\mathbf{k}_2\rangle.
$$

Here
$$
\langle\mathbf{k}_1|[H_0,\rho^{(1)}]|\mathbf{k}_2\rangle=(E_{\mathbf{k}_1}-
E_{\mathbf{k}_2})\langle\mathbf{k}_1|\rho^{(1)}|\mathbf{k}_2\rangle=$$$$=
(E_{\mathbf{k}_1}-E_{\mathbf{k}_2})\rho^{(1)}(\mathbf{k}_1-\mathbf{k}_2).
$$

Previous equality now we write in the form
$$
(\hbar\omega+i\hbar \nu-E_{\mathbf{k}_1}+E_{\mathbf{k}_2})
\rho^{(1)}(\mathbf{k}_1-\mathbf{k}_2)=
$$
$$
=\dfrac{e\hbar}{2mc}(f_{\mathbf{k}_1}-f_{\mathbf{k}_2})
(\mathbf{k}_1+\mathbf{k}_2)\mathbf{A}(\mathbf{k}_1-\mathbf{k}_2)+
i\hbar \nu \tilde{\rho}^{(1)}(\mathbf{k}_1-\mathbf{k}_2).
$$

Last summand in this equality we will replace according to the equation (1.9).
From the received equation it is found
$$
\rho^{(1)}(\mathbf{k}_1-\mathbf{k}_2)=
$$
$$=
\dfrac{e\hbar}{2mc}
\Xi(\mathbf{k}_1,\mathbf{k}_2)(f_{\mathbf{k}_1}-f_{\mathbf{k}_2})
(\mathbf{k}_1+\mathbf{k}_2)\mathbf{A}(\mathbf{k}_1-\mathbf{k}_2).
\eqno{(1.11)}
$$

Here
$$
\Xi(\mathbf{k}_1,\mathbf{k}_2)=
\dfrac{E_{\mathbf{k}_1}-E_{\mathbf{k}_2}-i\hbar \nu}
{(E_{\mathbf{k}_1}-E_{\mathbf{k}_2})(\hbar\omega+i\hbar \nu-
E_{\mathbf{k}_1}+E_{\mathbf{k}_2})}.
$$

In equation (1.11) we put $\mathbf{k}_1=\mathbf{k}$,
$\mathbf{k}_2=\mathbf{k}-\mathbf{q}$. Then
$$
\langle\mathbf{k}_1|\rho^{(1)}|\mathbf{k}_2\rangle=
\langle\mathbf{k}|\rho^{(1)}|\mathbf{k}-
\mathbf{q}\rangle=\rho^{(1)}(\mathbf{q})=
$$ \hspace{0.2cm}
$$
=-\dfrac{e\hbar}{mc}\Xi(\mathbf{k},\mathbf{q})
(f_{\mathbf{k}}-f_{\mathbf{k}-\mathbf{q}})
\mathbf{k}\mathbf{A}(\mathbf{q}).
\eqno{(1.12)}
$$\hspace{0.3cm}

Here
$$
\Xi(\mathbf{k},\mathbf{q})=
\dfrac{E_{\mathbf{k}}-E_{\mathbf{k-q}}-i\hbar \nu}
{(E_{\mathbf{k}}-E_{\mathbf{k-q}})(\hbar\omega+i\hbar \nu-
E_{\mathbf{k}}+E_{\mathbf{k-q}})}.
$$

\begin{center}
  \bf 2. Current density
\end{center}

Сurrent density ${\bf j}({\bf q})$ is defined as
$$
{\bf j}({\bf q},\omega)=e\int \dfrac{d{\bf k}}{8\pi^3m}\left\langle{\bf k}+
\frac{\mathbf{q}}{2}\left|({\bf p}-\frac{e}{c}{\bf A})
\rho+\rho ({\bf p}-\frac{e}{c}{\bf A} \big)\right|{\bf k}-
\frac{\mathbf{q}}{2}\right\rangle.
\eqno{(2.1)}
$$
After substitution (1.3) in integrand from (2.1), we have
$$
\left\langle{\bf k}+
\frac{\mathbf{q}}{2}\left|({\bf p}-\frac{e}{c}{\bf A})
\rho+\rho ({\bf p}-\frac{e}{c}{\bf A} \big)\right|{\bf k}-
\frac{\mathbf{q}}{2}\right\rangle=
$$
$$
=
\left\langle{\bf k}+
\frac{\mathbf{q}}{2}\left|{\bf p}\rho^{(1)}+\rho^{(1)}\mathbf{p}-\frac{e}{c}
({\bf A}\tilde{\rho}^{(0)}+\tilde{\rho}^{(0)}\mathbf{A})\right|{\bf k}-
\frac{\mathbf{q}}{2}\right\rangle.
$$
It is easy to see that
$$
\left\langle{\bf k}+\dfrac{\mathbf{q}}{2}\Big|{\bf p}\rho^{(1)}+
\rho^{(1)}\mathbf{p}\Big|{\bf k}-
\dfrac{\mathbf{q}}{2}\right\rangle
=2\hbar\mathbf{k}\rho^{(0)}(\mathbf{q}),
$$
$$
\left\langle{\bf k}+
\dfrac{\mathbf{q}}{2}\Big|{\bf A}\tilde{\rho}^{(0)}+
\tilde{\rho}^{(0)}\mathbf{A}\Big|{\bf k}-\dfrac{\mathbf{q}}{2}\right\rangle=
\mathbf{A}(\mathbf{q})\Big[\tilde{\rho}^{(0)}(\mathbf{k}+
\dfrac{\mathbf{q}}{2})+\tilde{\rho}^{(0)}(\mathbf{k}-
\dfrac{\mathbf{q}}{2})\Big].
$$

Hence, expression for current density has the form
$$
\mathbf{j}(\mathbf{q},\omega,\nu)=-\dfrac{e^2}{mc}\mathbf{A}(\mathbf{q})
\int \dfrac{d\mathbf{k}}{8\pi^3}\tilde{\rho}^{(0)}(\mathbf{k}+
\dfrac{\mathbf{q}}{2})-\dfrac{e^2}{mc}\mathbf{A}(\mathbf{q})
\int \dfrac{d\mathbf{k}}{8\pi^3}\tilde{\rho}^{(0)}(\mathbf{k}-
\dfrac{\mathbf{q}}{2})+
$$
$$
+e\hbar\int\dfrac{d\mathbf{k}}{4\pi^3m}
\left\langle\mathbf{k}+\dfrac{\mathbf{q}}{2}
\Big|\rho^{(1)}\Big|\mathbf{k}-\dfrac{\mathbf{q}}{2}\right\rangle.
$$

First two terms in this expression are equal each other
$$
\int \dfrac{d\mathbf{k}}{8\pi^3}\tilde{\rho}^{(0)}(\mathbf{k}+
\dfrac{\mathbf{q}}{2})=
\int \dfrac{d\mathbf{k}}{8\pi^3}\tilde{\rho}^{(0)}(\mathbf{k}-
\dfrac{\mathbf{q}}{2})=\dfrac{N}{2},
$$
where $N$ is the number density (concentration) of plasmas.

Therefore the current density is equal to
$$
{\bf j}({\bf q},\omega,\nu)=-\frac{e^2N}{mc}{\bf A}({\bf q})+
e\hbar\int \dfrac{d{\bf k}}{4\pi^3m}{\bf k}
\left\langle\mathbf{k}+\dfrac{\mathbf{q}}{2}\Big|\rho^{(1)}\Big|{\bf k}-
\frac{\mathbf{q}}{2}\right\rangle.
\eqno{(2.2)}
$$

The first term in (2.2) is the gauge density current.

By means of obvious replacement of variables the expression (2.2) is possible
to transform into the form
$$
{\bf j}({\bf q},\omega,\nu)=-\frac{e^2N}{mc}{\bf A}({\bf q})+
e\hbar\int \dfrac{d{\bf k}}{4\pi^3m}{\bf k}
\left\langle\mathbf{k}\Big|\rho^{(1)}\Big|{\bf k}-{\bf q}\right\rangle.
\eqno{(2.3)}
$$

In (2.3) the integrand is given by the equality (1.12).
Substituting (1.12) in (2.3), we obtain following expression for
 density current
$$
{\bf j}({\bf q},\omega,\nu)=-\frac{e^2N}{mc}{\bf A}({\bf q})-
$$
$$
-\dfrac{e^2\hbar^2}{m^2c}\int \dfrac{\mathbf{k}d\mathbf{k}}{4\pi^3}[\mathbf{k}
\mathbf{A(q)}]\Xi(\mathbf{k},\mathbf{q})(f_{\mathbf{k}}-f_{\mathbf{k-q}}).
\eqno{(2.4)}
$$

Let's direct an axis $x$ along the vector ${\bf q}$, and an axis $y$ along
the vector $ {\bf A} $. Then the previous expression (2.4) can be
overwritten in the form
$$
{ j}_y({\bf q},\omega,\nu)=-\frac{e^2N}{mc}{ A}({\bf q})-
\dfrac{e^2\hbar^2 A({\bf q})}{m^2c}\int \dfrac{d{\bf k}}{4\pi^3}{ k}_y^2
\;\Xi(\mathbf{k},\mathbf{q})(f_{\mathbf{k}}-f_{\mathbf{k-q}})
$$
and
$$
{ j}_x({\bf q},\omega,\nu)={ j}_z({\bf q},\omega,\nu)=0.
$$

Obviously that
$$
\int \dfrac{d{\bf k}}{4\pi^3}{ k}_y^2\;\Xi(\mathbf{k},\mathbf{q})
(f_{\mathbf{k}}-f_{\mathbf{k-q}})
=\int \dfrac{d{\bf k}}{4\pi^3}{ k}_z^2\;\Xi(\mathbf{k},\mathbf{q})
(f_{\mathbf{k}}-f_{\mathbf{k-q}}).
$$
Therefore
$$
\int \dfrac{d{\bf k}}{4\pi^3}{ k}_y^2\;
\Xi(\mathbf{k},\mathbf{q})(f_{\mathbf{k}}-f_{\mathbf{k-q}})
=\dfrac{1}{2}\int \dfrac{d{\bf k}}{4\pi^3}({ k}_y^2+{
k}_z^2)\;\Xi(\mathbf{k},\mathbf{q})(f_{\mathbf{k}}-f_{\mathbf{k-q}})=
$$
$$
=
\dfrac{1}{2}\int \dfrac{d{\bf k}}{4\pi^3}({\bf k}^2-{
k}_x^2)\;\Xi(\mathbf{k},\mathbf{q})(f_{\mathbf{k}}-f_{\mathbf{k-q}}).
$$
From here we conclude, that expression for density current is possible
to present in the following invariant form
$$\boxed{
{\bf j}({\bf q},\omega,\nu)=-{\bf A}({\bf q})\Big[\frac{Ne^2}{mc}+
\dfrac{e^2\hbar^2}{8\pi^3m^2c}\int d{\bf k}
\mathbf{k}^2_\perp
\Xi(\mathbf{k},\mathbf{q})(f_{\mathbf{k}}-f_{\mathbf{k-q}})\Big]}.
\eqno{(2.5)}
$$

Here
$$
\mathbf{k}^2_\perp={\bf k}^2-\Big(\dfrac{{\bf k}{\bf q}}{q}\Big)^2.
$$

Considering partial-fraction decomposition
$$
\Xi(\mathbf{k},\mathbf{q})=\dfrac{1}{E_{\mathbf{k}}-E_{\mathbf{k-q}}}+
\dfrac{\hbar \omega}{(E_{\mathbf{k}}-E_{\mathbf{k-q}})
[ E_{{\bf k}}-E_{{\bf k-q}}-\hbar( \omega+i\nu)]},
$$
we will present current density in the form
$$
{\bf j}({\bf q},\omega,\nu)=-{\bf A}({\bf q})\Big[\frac{Ne^2}{mc}+
\dfrac{e^2\hbar^2}{8\pi^3m^2c}\int \dfrac{f_{{\bf k}}-f_{{\bf k-q}}}
{E_{{\bf k}}-E_{{\bf k-q}}}\mathbf{k}^2_\perp d{\bf k}+
$$\medskip
$$
+\dfrac{e^2\hbar^3\omega}{8\pi^3m^2c}\int
 \dfrac{(f_{{\bf k}}-f_{{\bf k-q}})\mathbf{k}^2_\perp d{\bf k}}{(E_{{\bf k}}-E_{{\bf k-q}})
 [ E_{{\bf k}}-E_{{\bf k-q}}-\hbar( \omega+i\nu)]}
\Big].
\eqno{(2.6)}
$$\medskip

First two terms in the previous equality (2.6) do not depend on the frequency
$\omega$
and these terms defined by dissipative properties of a material.
Dissipative properties of material defined by collision frequency $\nu$.
These terms are the universal parameters defining {\it Landau diamagnetism}.

\begin{center}
\bf 3. Transversal conductivity and permeability
\end{center}

Considering the relation between vector potential with intensity of an
electromagnetic field,
and also relation of density of a current with electric field, on the basis
of the previous equality (2.5) it is received following expression of an
invariant form for the transversal electric conductivity
$$
\sigma_{tr}(\mathbf{q},\omega,\nu)=
\dfrac{ie^2N}{m\omega}+\dfrac{ie^2\hbar^2}{8\pi^3m^2\omega}
\int \Xi(\mathbf{k},\mathbf{q})(f_{\mathbf{k}}-f_{\mathbf{k-q}})
\mathbf{k}_\perp^2d\mathbf{k}.
\eqno{(3.1)}
$$

Allocating in (3.1) static conductivity $\sigma_0=e^2N/m\nu $,
let's overwrtite (3.1) in the form
$$
\dfrac{\sigma_{tr}(\mathbf{q},\omega,\nu)}{\sigma_0}=\dfrac{i \nu}{\omega}
\Big[1+\dfrac{\hbar^2}{8\pi^3mN}\int \Xi(\mathbf{k,q})(f_{\mathbf{k}}-
f_{\mathbf{k-q}})\mathbf{k}_\perp^2 d\mathbf{k}\Big].
$$

Let's take advantage of definition of transversal dielectric permeability
$$
\varepsilon_{tr}(\mathbf{q},\omega,\nu)=
1+\dfrac{4\pi i}{\omega}\sigma_{tr}(\mathbf{q},\omega,\nu).
\eqno{(3.2)}
$$

Taking into account (3.1) and equality (3.2) we will write expression
for the transversal dielectric permeability
$$
\varepsilon_{tr}(\mathbf{q},\omega,\nu)=1-\dfrac{\omega_p^2}{\omega^2}
\Big[1+\dfrac{\hbar^2}{8\pi^3mN}\int \Xi(\mathbf{k,q})(f_{\mathbf{k}}-
f_{\mathbf{k-q}})\mathbf{k}_\perp^2 d\mathbf{k}\Big].
\eqno{(3.3)}
$$

Here $\omega_p$ is the plasma (Langmuir) frequency,
$\omega_p^2=4\pi e^2N/m$.

From equality (3.3) it is visible, that one of equalities named a rule
$f$-sums is carried out (see, for example, \cite{Dressel}, \cite{Pains} and
\cite{Martin}) for the transversal dielectric permeability. This rule is
expressed by the formula (4.200) from the monography \cite{Pains}
$$\boxed{
\int\limits_{-\infty}^{\infty}\varepsilon_{tr}(\mathbf{q},\omega,\nu)\omega
d\omega=\pi \omega_p^2}.
$$

Let's decompose expression $ \Xi(\mathbf{k},\mathbf{q})$ from the subintegral
expressions from (3.1) on partial fractions
$$
\Xi(\mathbf{k},\mathbf{q})=\dfrac{E_{\mathbf{k}}-E_{\mathbf{k-q}}-i\hbar \nu}
{(E_{\mathbf{k}}-E_{\mathbf{k-q}})[E_{\mathbf{k}}-E_{\mathbf{k-q}}-\hbar
(\omega+i\nu)]}=
$$
$$
=\dfrac{i \nu}{\omega+i \nu}\dfrac{1}
{E_{\mathbf{k}}-E_{\mathbf{k-q}}}+\dfrac{\omega}{\omega+i \nu}\dfrac{1}
{E_{\mathbf{k}}-E_{\mathbf{k-q}}-\hbar (\omega+i\nu)}.
$$

Hence, for transversal electric conductivity and dielectric
permeability it is had following obvious representations
$$
\dfrac{\sigma_{tr}(\mathbf{q},\omega,\nu)}{\sigma_0}=
\dfrac{i \nu}{\omega}\Bigg[1+\dfrac{\hbar^2}{8\pi^3mN(\omega+i \nu)}
\Bigg(i \nu\int \dfrac{f_{{\bf k}}-f_{{\bf k-q}}}{E_{{\bf k}}-E_{{\bf k-q}}}
\mathbf{k}^2_\perp d{\bf k}+
$$
$$
+\omega\int
\dfrac{(f_{{\bf k}}-f_{{\bf k-q}})\mathbf{k}^2_\perp d{\bf k}}
{E_{{\bf k}}-E_{{\bf k-q}}-\hbar(\omega+i\nu)}\Bigg)\Bigg]
\eqno{(3.4)}
$$
and
$$
\varepsilon_{tr}(\mathbf{q},\omega,\nu)=1-\dfrac{\omega_p^2}{\omega^2}
\Bigg[1+\dfrac{\hbar^2}{8\pi^3mN(\omega+i \nu)}
\Bigg(i \nu\int \dfrac{f_{{\bf k}}-f_{{\bf k-q}}}{E_{{\bf k}}-E_{{\bf k-q}}}
\mathbf{k}^2_\perp d{\bf k}+
$$
$$
+\omega\int
\dfrac{(f_{{\bf k}}-f_{{\bf k-q}})\mathbf{k}^2_\perp d{\bf k}}
{E_{{\bf k}}-E_{{\bf k-q}}-\hbar(\omega+i\nu)}\Bigg)\Bigg]
\eqno{(3.5)}
$$

If to enter designations
$$
J_\nu=\dfrac{\hbar^2}{8\pi^3mN}
\int \dfrac{f_{{\bf k}}-f_{{\bf k-q}}}{E_{{\bf k}}-E_{{\bf k-q}}}
\mathbf{k}^2_\perp d{\bf k}
$$
and
$$
J_\omega=\dfrac{\hbar^2}{8\pi^3mN}\int
\dfrac{(f_{{\bf k}}-f_{{\bf k-q}})\mathbf{k}^2_\perp d{\bf k}}
{E_{{\bf k}}-E_{{\bf k-q}}-\hbar(\omega+i\nu)},
$$
then expressions (3.4) and (3.5) become simpler
$$\boxed{
\dfrac{\sigma_{tr}(\mathbf{q},\omega,\nu)}{\sigma_0}=
\dfrac{i \nu}{\omega}\Big(1+\dfrac{\omega J_\omega+i \nu J_\nu}{\omega+i \nu}
\Big)}
\eqno{(3.6)}
$$
and
$$\boxed{
\varepsilon_{tr}(\mathbf{q},\omega,\nu)=
1-\dfrac{\omega_p^2}{\omega^2}
\Big(1+\dfrac{\omega J_\omega+i \nu J_\nu}{\omega+i \nu}
\Big)}.
\eqno{(3.7)}
$$

Integrals $J_\omega $ and $J_\nu $ can be transformed to the following form
$$
J_\omega=\dfrac{\hbar^2}{8\pi^3mN}\int \dfrac{[2E_\mathbf{k}-(E_\mathbf{k-q}-
E_\mathbf{k+q})]f_\mathbf{k}\mathbf{k}_\perp^2d\mathbf{k}}
{[E_\mathbf{k}-E_\mathbf{k-q}-\hbar(\omega+i \nu)]
[E_\mathbf{k}-E_\mathbf{k+q}+\hbar(\omega+i \nu)]}
$$
and
$$
J_\nu=\dfrac{\hbar^2}{8\pi^3mN}\int \dfrac{[2E_\mathbf{k}-(E_\mathbf{k-q}-
E_\mathbf{k+q})]f_\mathbf{k}\mathbf{k}_\perp^2d\mathbf{k}}
{(E_\mathbf{k}-E_\mathbf{k-q})(E_\mathbf{k}-E_\mathbf{k+q})}.
$$

In integrals from expressions (3.6) and (3.7) we have
$$
\mathbf{k}^2_\perp d\mathbf{k}=(k^2-k_x^2)d^3k=(k_y^2+k_z^2)d^3k.
$$

Instead of  vector $ \mathbf{k} $ we will enter the dimensionless vector
$\mathbf{K}$ by the following equality
$\mathbf{K} = \dfrac {\mathbf {k}}{k_F} $, \quad
$k_F =\dfrac {p_F}{\hbar} $, where $k_F $ is the wave number of Fermi,
$p_F=mv_F $ is the electron momentum on Fermi's surface, $v_F $ is the
electron  velocity on Fermi's surface.

Then
$$
\mathbf{k}_\perp^2d\mathbf{k}=k_F^5(K^2-K_x^2)d^3K=k_F^5(K_y^2+K_z^2)d^3K=
k_F^5K_\perp^2d^3K,
$$
where
$$
K_\perp^2=K^2-K_x^2=K_y^2+K_z^2.
$$
Energy $E_{\mathbf {k}}$ we will express through Fermi's energy
$E_F =\dfrac{mv_F^2}{2}$. We have
$$
E_{\mathbf{k}}=\dfrac{\hbar^2\mathbf{k}^2}{2m}=\dfrac{\hbar^2k_F^2}{2m}
\mathbf{K}^2=\dfrac{p_F^2}{2m}\mathbf{K}^2=E_F\mathbf{K}^2\equiv E_{\mathbf{K}}.
$$

In the same way we receive
$$
E_{\mathbf{k-q}}=\dfrac{\hbar^2(k_F\mathbf{K}-\mathbf{q})^2}{2m}=
\dfrac{\hbar^2k_F^2}{2m}\Big(\mathbf{K}-\dfrac{\mathbf{q}}{k_F}\Big)^2.
$$
Further we will designate a wave vector $\mathbf{q}$ through $\mathbf{k}$,
and we will enter dimensionless wave vector
$\mathbf{q} =\dfrac{\mathbf{k}}{k_F}$. Then
$$
E_{\mathbf{k-q}}=\dfrac{\hbar^2k_F^2}{2m}\Big(\mathbf{K}-\mathbf{q}\Big)^2=
E_F(\mathbf{K-q})^2=E_\mathbf{K-q}.
$$

Hence, taking into account last equalities of the formula for integrals
$J_\omega $ and $J_\nu $ it is possible to overwrite in the form
$$
J_\nu=\dfrac{\hbar^2k_F^5}{8\pi^3mN}\int \dfrac{f_\mathbf{K}-f_\mathbf{K-q}}
{E_\mathbf{K}-E_\mathbf{K-q}}K_\perp^2d^3K
$$
and
$$
J_\omega=\dfrac{\hbar^2k_F^5}{8\pi^3mN}\int \dfrac{f_\mathbf{K}-f_\mathbf{K-q}}
{E_\mathbf{K}-E_\mathbf{K-q}-\hbar(\omega+i \nu)}K_\perp^2d^3K.
$$

Here
$$
f_\mathbf{K}=\Big[1+\exp\dfrac{E_\mathbf{K}-\mu}{k_BT}\Big]^{-1}.
$$

\begin{center}
\bf 4. Degenerate plasma
\end{center}

Let's consider further a case of degenerate plasmas. Then we have
$$
\Big(\dfrac{mv_F}{\hbar}\Big)^3 \equiv \Big(\dfrac{p_F}{\hbar}\Big)^3\equiv
k_F^3=3\pi^2N.
$$

Absolute distribution of Fermi---Dirac $f_\mathbf{K}$
for degenerate plasmas passes in Fermi's distribution
$$
f_\mathbf{K}=\Theta_\mathbf{K}\equiv\Theta(E_F-E_{\mathbf{K}})=
\Theta(1-\mathbf{K}^2).
$$
Here $\Theta(x)$ is the Heaviside function,
$$
\Theta(x)=\left\{\begin{array}{c}
                   1,\qquad x>0, \\
                   0,\qquad x<0.
                 \end{array}\right.
$$
In the same way we receive
$$
\Theta_\mathbf{K-q}=\Theta(E_F-E_{\mathbf{K-q}})=
\Theta[E_F-E_F(\mathbf{K-q})^2]=\Theta[1-(\mathbf{K-q})^2].
$$

Taking into account these equalities these expression for integrals $J_\omega $
and $J_\nu $ will be transformed to the form
$$
J_\omega=\dfrac{3E_F}{4\pi}\int\dfrac{\Theta_\mathbf{K}-E_\mathbf{K-q}}
{E_\mathbf{K}-E_\mathbf{K-q}-\hbar(\omega+i \nu)}K_\perp^2d^3K
$$
and
$$
J_\nu=\dfrac{3E_F}{4\pi}\int\dfrac{\Theta_\mathbf{K}-E_\mathbf{K-q}}
{E_\mathbf{K}-E_\mathbf{K-q}}K_\perp^2d^3K.
$$

Calculation of transversal electric conductivity and dielectric
permeability we will spend under formulas (3.6) and (3.7).
Let's calculate integrals $J_\omega $ and $J_\nu $. We will notice that
$$
E_{\mathbf{K}}-E_{\mathbf{K-q}}=E_F\mathbf{K}^2-E_F(\mathbf{K-q})^2=
E_F[2K_x q-q^2]=
$$
$$
=2qE_F(K_x-\dfrac{q}{2})=mv_F^2q(K_x-\dfrac{q}{2})=
\hbar v_Fk(K_x-\dfrac{q}{2}).
$$

Besides,
$$
E_{\mathbf{K}}-E_{\mathbf{K-q}}-\hbar (\omega+i\nu)=
mv_F^2q\Big(K_x-\dfrac{q}{2}-\dfrac{\omega+i \nu}{v_Fk}\Big)=$$$$=
\hbar v_Fk\Big(K_x-\dfrac{q}{2}-\dfrac{\omega+i \nu}{v_Fk_F q}\Big)=
\hbar v_F k\Big(K_x-\dfrac{q}{2}-\dfrac{z}{q}\Big),
$$
where
$$
z=x+iy,\quad x=\dfrac{\omega}{k_Fv_F}=\dfrac{\hbar \omega}{2E_F},\quad
y=\dfrac{\nu}{k_Fv_F}=\dfrac{\hbar\nu}{2E_F},
\quad q=\dfrac{k}{k_F}.
$$

By means of the previous equalities for integrals $J_\omega $ and $J_\nu $
we receive the following expressions
$$
J_\omega=\dfrac{3}{8\pi q}\int \dfrac{\Theta_\mathbf{K}-\Theta_\mathbf{K-q}}
{K_x-q/2-z/q}K_\perp^2d^3{K}
$$
and
$$
J_\nu=\dfrac{3}{8\pi q}\int \dfrac{\Theta_\mathbf{K}-\Theta_\mathbf{K-q}}
{K_x-q/2}K_\perp^2d^3{K}.
$$

Let's transform the first integral
$$
J_\omega=\dfrac{3}{8\pi q}\int \dfrac{\Theta(1-\mathbf{K}^2)-\Theta[1-
\mathbf{(K-q)}]^2}{K_x-z/q-q/2}K_\perp^2d^3{K}=
$$
$$
=\dfrac{3}{8\pi q}\int\Big[\dfrac{1}{K_x-z/q-q/2}-\dfrac{1}
{K_x-z/q+q/2}\Big]\Theta(1-\mathbf{K}^2)K_\perp^2d^3{K}=
$$
$$
=\dfrac{3}{8 \pi}\int \dfrac{\Theta(1-\mathbf{K}^2)K_\perp^2d^3K}
{(K_x-z/q)^2-q^2/4}d^3K=\dfrac{3}{16}\int\limits_{-1}^{1}
\dfrac{(1-t^2)^2dt}{(t-z/q)^2-q^2/4}.
$$

Arguing similarly we will present the second integral in the form
$$
J_\nu=\dfrac{3}{8\pi q}\int \dfrac{\Theta_\mathbf{K}-\Theta_\mathbf{K-q}}
{K_x-q/2}K_\perp^2d^3{K}=
$$
$$
=\dfrac{3}{16}\int\limits_{-1}^{1}
\dfrac{(1-t^2)^2dt}{t^2-q^2/4}=\dfrac{3}{8}\int\limits_{0}^{1}
\dfrac{(1-t^2)^2dt}{t^2-q^2/4}.
$$

Let's return to formulas (3.6) and (3.7) and we will present them to the
dimensionless parametres
$$\boxed{
\dfrac{\sigma_0(q,x,y)}{\sigma_0}=\dfrac{iy}{x}\Big(1+\dfrac{xJ_\omega+
iyJ_\nu}{x+iy}\Big)}
\eqno{(4.1)}
$$
and
$$\boxed{
\varepsilon_{tr}(q,x,y)=1-\dfrac{x_p^2}{x^2}\Big(1+\dfrac{xJ_\omega+
iyJ_\nu}{x+iy}\Big)}.
\eqno{(4.2)}
$$
Here $x_p$ is the dimensionless plasma (Langmuir) frequency,
$$x_p=\dfrac{\omega_p}{k_Fv_F}=\dfrac{\hbar \omega_p}{2E_F}.
$$

Let's present formulas (4.1) and (4.2) in an explicit form
$$
\dfrac{\sigma_{tr}}{\sigma_0}=\dfrac{iy}{x}\Bigg[1+\dfrac{3}{8(x+iy)}
\Bigg(x\int\limits_{0}^{1}\dfrac{(1-t^2)^2dt}{t^2-q^2/4}+
\dfrac{iy}{2}\int\limits_{-1}^{1}
\dfrac{(1-t^2)^2dt}{(t-z/q)^2-q^2/4}\Bigg)\Bigg]
\eqno{(4.3)}
$$
and
$$
\varepsilon_{tr}=1-\dfrac{x_p^2}{x^2}\Bigg[1+\dfrac{3}{8(x+iy)}
\Bigg(x\int\limits_{0}^{1}\dfrac{(1-t^2)^2dt}{t^2-q^2/4}+
\dfrac{iy}{2}\int\limits_{-1}^{1}
\dfrac{(1-t^2)^2dt}{(t-z/q)^2-q^2/4}\Bigg)\Bigg]
$$

For comparison we will present Lindhard's formula \cite{Lin} in our designations
as follows
$$
\dfrac{\sigma_{tr}^{Lin}}{\sigma_0}=\dfrac{iy}{x}+\dfrac{3iy}{16x}
\int\limits_{-1}^{1}
\dfrac{(1-t^2)^2dt}{(t-z/q)^2-q^2/4}.
\eqno{(4.4)}
$$

From equalities (4.3) and (4.4) it is visible, that imaginary parts
transversal electric conductivities from the present work and from
Lindhard's work at $y\to 0$ coincide
$$
\lim\limits_{y\to 0}\Im \sigma_{tr}(\mathbf{q},\omega,\nu)=
\Im \sigma_{tr}^{Lin}(\mathbf{q},\omega,\nu).
$$

The integrals entering in (4.3), are easily calculated
$$
T_0(q)=\int\limits_{0}^{1}\dfrac{(1-t^2)^2dt}{t^2-q^2/4}=-\dfrac{5}{3}+
\dfrac{q^2}{4}+\dfrac{(q^2-4)^2}{16q}\ln\dfrac{2-q}{2+q}=
$$
$$
=-\dfrac{5}{3}+\dfrac{k^2}{4k_F^2}+\dfrac{(k^2-4k_F^2)^2}{16kk_F^3}
\ln\dfrac{2k_F-k}{2k_F+k},
\eqno{(4.5)}
$$
$$
T_1(q,z)=\int\limits_{-1}^{1}\dfrac{(1-t^2)^2dt}{(t-z/q)^2-q^2/4}=
-\dfrac{10}{3}+\dfrac{4z^2}{q^2}+\dfrac{q^2}{2}+
\dfrac{1}{q}\Big[\Big(1-\dfrac{z^2}{q^2}\Big)^2+\dfrac{q^4}{16}-$$
$$
-\dfrac{q^2}{2}+\dfrac{3z^2}{2}\Big]\ln\dfrac{(1-q/2)^2-z^2/q^2}
{(1+q/2)^2-z^2/q^2}-
\dfrac{zq}{2}\Big[1+\dfrac{4}{q^2}\Big(1-\dfrac{z^2}{q^2}\Big)\Big]
\ln\dfrac{(1-z/q)^2-q^2/4}{(1+z/q)^2-q^2/4}.
\eqno{(4.6)}
$$

Hence, transversal electric conductivity of quantum plasma it is calculated
under the formula
$$\boxed{
\dfrac{\sigma_{tr}}{\sigma_0}=iy
\dfrac{3xT_1(q,z)+6iyT_0(q)}{16x(x+iy)}},
\eqno{(4.7)}
$$
where $T_0(q)$ and $T_1(q,z)$ are given accordingly by formulas (4.5) and
(4.6).

Similarly, the transversal dielectric permeability is calculated under
the formula
$$\boxed{
\varepsilon_{tr}=
1-\dfrac{x_p^2}{x^2}\cdot\dfrac{3xT_1(q,z)+6iyT_0(q)}{16(x+iy)}}.
$$

Let's present the formula for transversal electric conductivity
in an explicit form
$$
\dfrac{\sigma_{tr}}{\sigma_0}=\dfrac{iy}{x}\Bigg\{1+\dfrac{3iy}{x}
\Big[-\dfrac{5}{3}+
\dfrac{q^2}{4}+\dfrac{(q^2-4)^2}{16q}\ln\dfrac{2-q}{2+q}\Big]+
\dfrac{3x}{16z}\bigg[
-\dfrac{10}{3}+\dfrac{4z^2}{q^2}+
$$
$$
+\dfrac{q^2}{2}+\dfrac{1}{q}\Big(\Big(1-\dfrac{z^2}{q^2}\Big)^2+\dfrac{q^4}{16}-
\dfrac{q^2}{2}+\dfrac{3z^2}{2}\Big)\ln\dfrac{(1-q/2)^2-z^2/q^2}
{(1+q/2)^2-z^2/q^2}-
$$
$$
-\dfrac{zq}{2}\Big[1+\dfrac{4}{q^2}\Big(1-\dfrac{z^2}{q^2}\Big)\Big]
\ln\dfrac{(1-z/q)^2-q^2/4}{(1+z/q)^2-q^2/4}
\bigg]\Bigg\}.
$$

\begin{center}
  \bf 5. Specal cases of electric conductivity
\end{center}

We investigate special cases of electric conductivity.
We take the formula (3.6) and we will transform it to the form
$$
\dfrac{\sigma_{tr}}{\sigma_0}=\dfrac{i \nu}{\omega}-\dfrac{3\nu^2}
{8\pi\omega(\omega+i \nu)q}
\int\dfrac{\Theta_\mathbf{K}-\Theta_\mathbf{K-q}}{K_x-q/2}K_\perp^2d^3K+
$$
$$
+\dfrac{3i \nu}{8\pi(\omega+i \nu)q}\int
\dfrac{\Theta_\mathbf{K}-\Theta_\mathbf{K-q}}{K_x-z/q-q/2}K_\perp^2d^3K.
\eqno{(5.1)}
$$

Subintegral expressions in these integrals contain function
$
\varphi(\mathbf{q})=\Theta(1-K^2)-\Theta(1-(\mathbf{K-q})^2).
$
In linear approximation we have
$
\varphi(\mathbf{q})=-2\delta(1-K^2)K_xq=-\delta(1-K)K_xq.
$
Now expression (5.1) becomes simpler
$$
\dfrac{\sigma_{tr}}{\sigma_0}=\dfrac{i \nu}{\omega}\Bigg[1-\dfrac{3i\nu}
{8\pi(\omega+i \nu)}
\int\dfrac{K_x\delta(1-K)K_\perp^2d^3K}{K_x-q/2}-
$$
$$
-\dfrac{3\omega}{8\pi(\omega+i \nu)}\int
\dfrac{K_x\delta(1-K)K_\perp^2d^3K}
{K_x-z/q-q/2}\Bigg].
\eqno{(5.2)}
$$
We notice that
$$
\dfrac{3}{8\pi}\int \delta(1-K)K_\perp^2d^3K=1.
\eqno{(5.3)}
$$

Hence, expression (5.2) by means of (5.3) can be transformed
to the following form
$$
\dfrac{\sigma_{tr}}{\sigma_0}=\dfrac{i \nu}{\omega}\Bigg[
\dfrac{\omega}{\omega+i \nu}-\dfrac{3i\nu q}
{16\pi(\omega+i \nu)}
\int\dfrac{\delta(1-K)K_\perp^2d^3K}{K_x-q/2}-
$$
$$
-\dfrac{3\omega}{8\pi(\omega+i \nu)}\int
\dfrac{K_x\delta(1-K)K_\perp^2d^3K}
{K_x-z/q-q/2}\Bigg].
\eqno{(5.4)}
$$

Now from (5.4) it is clear, that at $q\to 0$
$
\sigma_{tr}=\dfrac{i \nu}{\omega+i \nu}\sigma_0,
$
and at $\omega=0$ we have exactly the static conductivity:
$\sigma_{tr}=\sigma_0$.

Let's show now that at small $q$ the expression (5.4) leads to known
exp\-res\-sion for conductivity of degenerate Fermi plasma. Really, we
will notice, that at small $q$ the first integral from (5.4) is
proportional to $q^2$. In denominator of the second integral we will neglect
the term $q/2$, because $q/2 \ll |z|/q $. Now as result for small $q$
we have
$$
\dfrac{\sigma_{tr}}{\sigma_0}=\dfrac{i \nu}{\omega + i
\nu}\Bigg[1-\dfrac{3}{8\pi}\int\dfrac{\delta(1-K)K_x}{K_x-z/q}K_\perp^2d^3K
\Bigg].
\eqno{(5.5)}
$$

Using again equality (5.3), we come to following exoression
$$
\dfrac{\sigma_{tr}}{\sigma_0}=-\dfrac{3y}{8\pi q}\int \dfrac{\delta(1-K)
K_\perp^2}{K_x-q/2}d^3K,
$$
which leads to known expression for the electric conductivity in degenerate
plasma
$$
\sigma_{tr}^{\rm classic}=-\dfrac{3iy\sigma_0}{4q}
\int\limits_{-1}^{1}\dfrac{1-\mu^2}{\mu-z}d\mu=
\dfrac{3iy\sigma_0}{4q}\Big[\dfrac{2z}{q}+(z^2-q^2)\ln\dfrac{z-q}{z+q}\Big].
\eqno{(5.6)}
$$
\begin{figure}[h]
\begin{center}
\includegraphics[width=17.0cm, height=8cm]{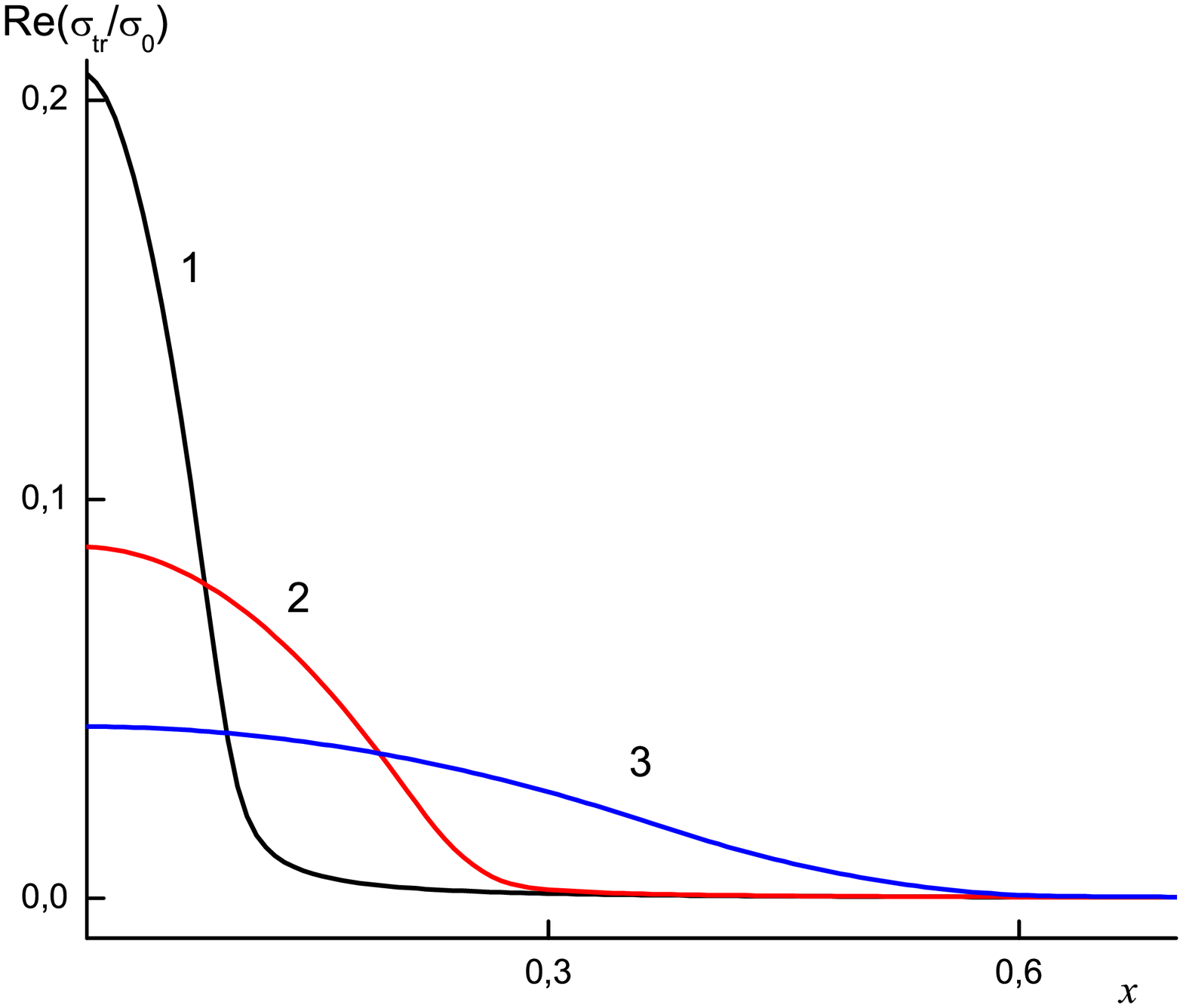}
\end{center}
\begin{center}
{Fig. 1.}
\end{center}
\end{figure}~
On figs. 1 and 2 we will present dependence of the real and imaginary parts
of the transversal electric conductivity deduced
in the present work, from dimensionless frequency of oscillations $x$.
Curves $1,2,3$ answer to values of dimensionless wave number $q=0.1,0.25,0.5$.
From these drawings it is visible, that the real part is monotonously decreasing
function, and the imaginary part has a maximum which
decreases with growth of wave number.
With growth of frequency of oscillations of plasma and
real, and imaginary parts decrease to zero.

On figs. 3--5 module comparison (fig. 3) is given, and also
real (fig. 4) and imaginary (fig. 5) parts of the transversal
conductivity (curves 1), conductivity on Linhard (curves 2) and
classical conductivity (curves 3).

On fig. 3 comparison of conductivity modules depending on the dimen\-si\-on\-less
frequencies of oscillations of plasma is spent, by this $y=0.1, q=1$.
From graps it is visible,
that at $x\to 0$ conductivity on Lindhard coincides with conductivity
of quantum plasma, and at $x\to \infty $ all three conductivities coincide.
On figs. 4 and 5 studying of the real and imaginary parts of
con\-duc\-ti\-vi\-ti\-es
depending on dimensionless wave number is spent, at this
$y=0.01, x=0.1$. From these graphs it is visible, that at small values
of collision frequencies of plasma particles conductivity on Lindhard
practically coincides with conductivity of plasma at small values
wave number.

\begin{center}
\bf 6. Conclusion
\end{center}

In the present work formulas for electric conductivity and dielect\-ric
per\-me\-abi\-lity of quantum collisinal plasma in Mermin' approach are deduced.

For this purpose the kinetic equation with integral of collisions in form
of relaxation model in momentum space is used.
Various special cases are investigated.
The case of degenerate Fermi plasmas is allocated and investi\-ga\-ted.
Graphic comparison of conductivity from the present work with
con\-duc\-ti\-vi\-ty under Linhard and with classical conductivity is carry out.

\begin{figure}[h]
\begin{center}
\includegraphics[width=17.0cm, height=10cm]{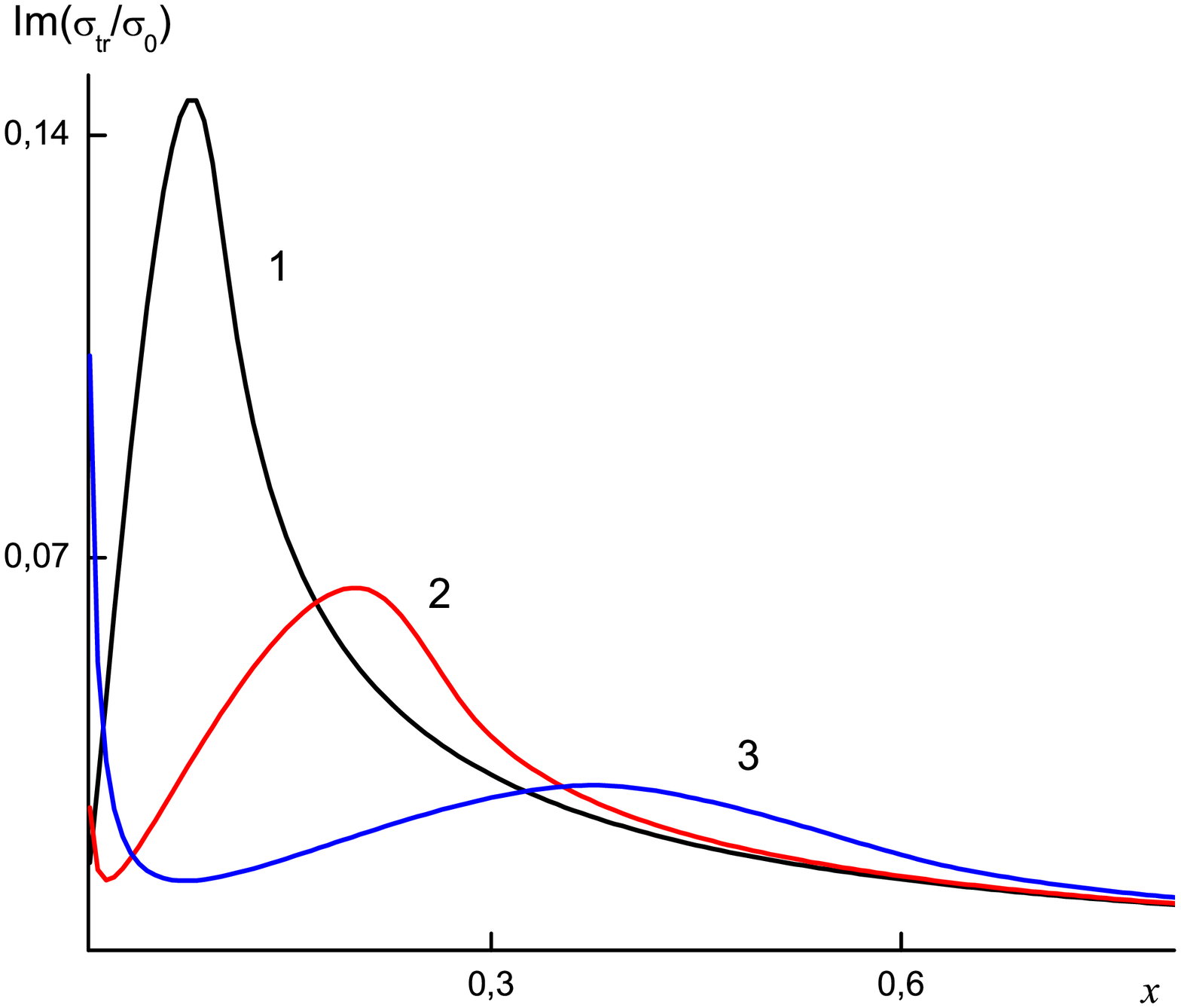}
\end{center}
\begin{center}
{Fig. 2.}
\end{center}
\begin{center}
\includegraphics[width=17.0cm, height=10cm]{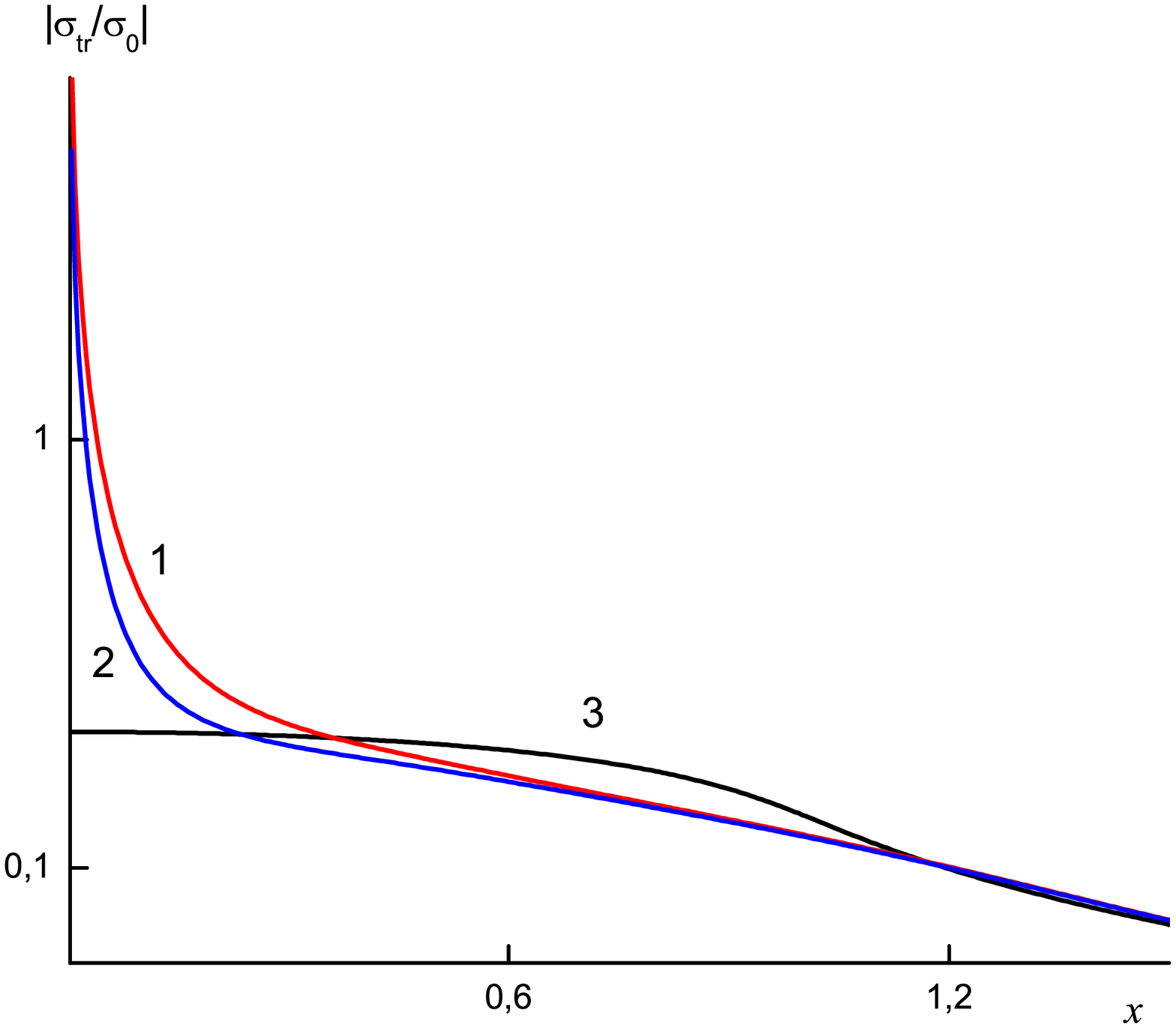}
\end{center}
\begin{center}
{Fig. 3.}
\end{center}
\end{figure}

\begin{figure}[h]
\begin{center}
\includegraphics[width=17.0cm, height=9cm]{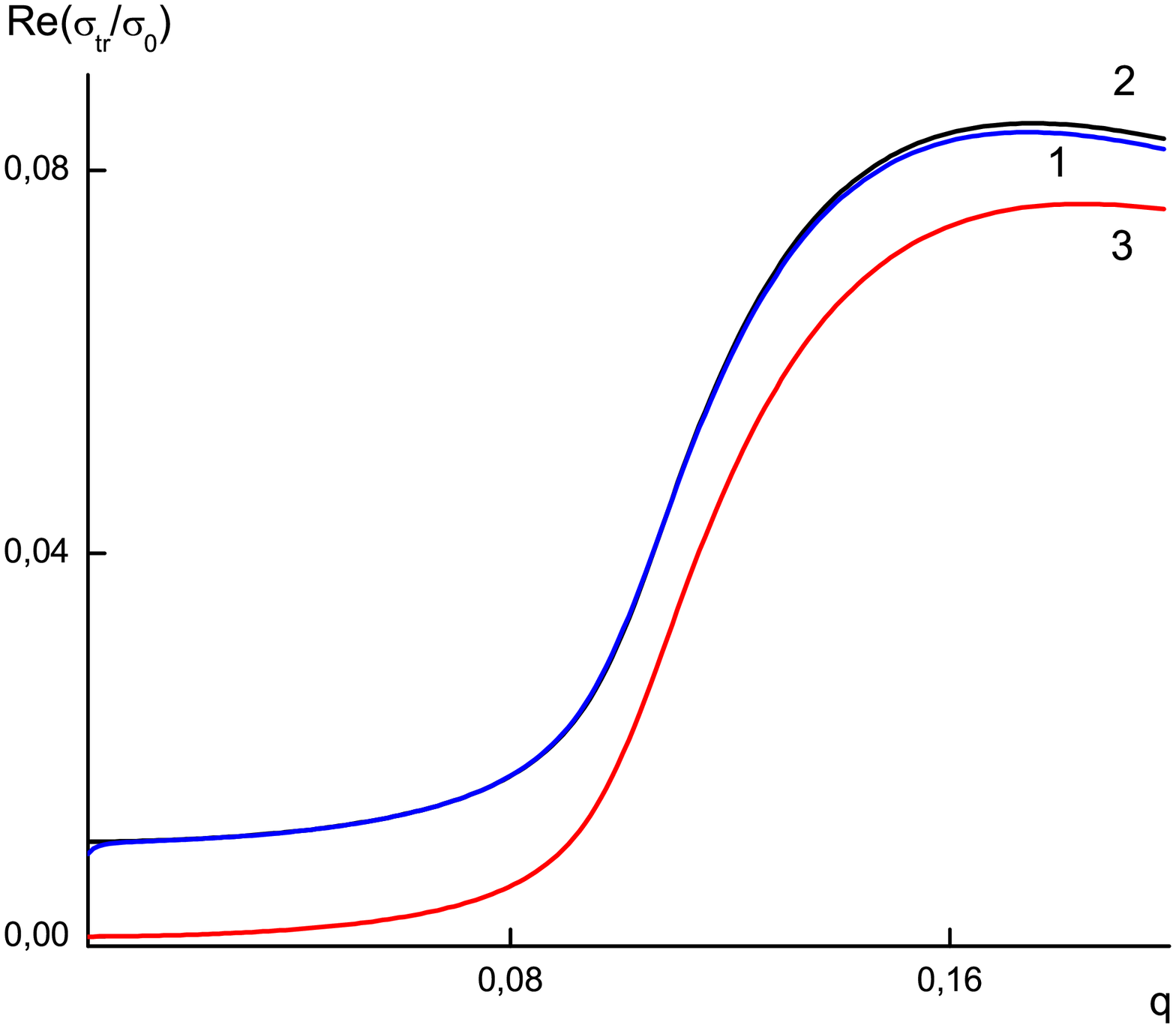}
\end{center}
\begin{center}
{Fig. 4.}
\end{center}
\end{figure}
\begin{figure}[h]
\begin{center}
\includegraphics[width=17.0cm, height=9cm]{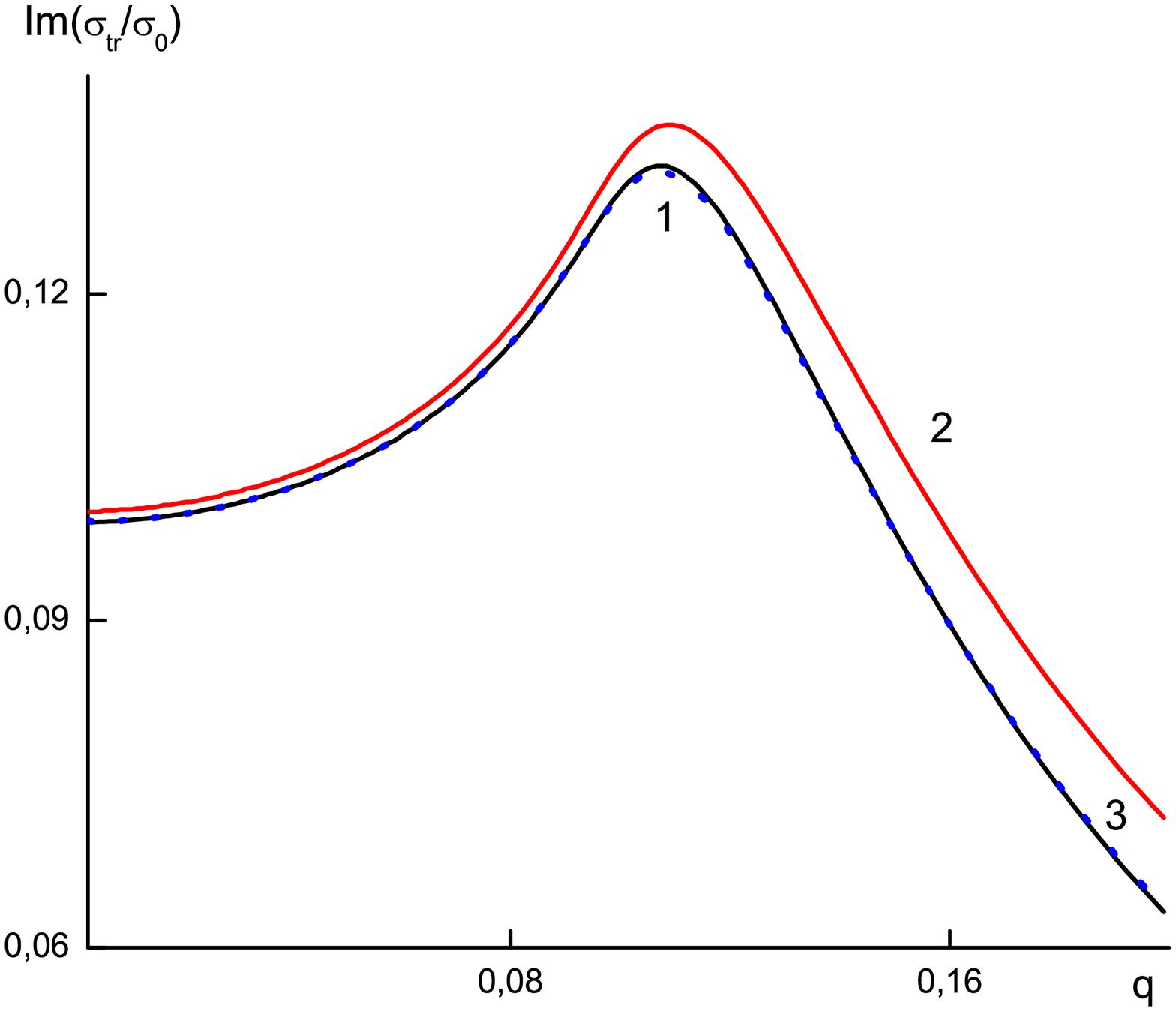}
\end{center}
\begin{center}
{Fig. 5.}
\end{center}
\end{figure}

\end{document}